\documentclass[aps,superscriptaddress,showpacs,nofootinbib,eqsecnum]{revtex4}

\begin{document}

\title{Evaluating Critical Exponents in the Optimized Perturbation
Theory}

\author{Marcus Benghi Pinto}
\email{marcus@fsc.ufsc.br}
\affiliation{Departamento de
F\'{\i}sica, Universidade Federal de Santa Catarina, 88040-900
Florian\'{o}polis, SC, Brazil}

\author{Rudnei O. Ramos}
\email{rudnei@uerj.br}
\affiliation{Departamento de F\'{\i}sica
Te\'orica, Universidade do Estado do Rio de Janeiro, 20550-013 Rio
de Janeiro, RJ, Brazil}

\author{Paulo J. Sena}
\email{psena@bon.matrix.com.br}
\affiliation{Unisul, Avenida
Jos\'{e} Ac\'{a}cio Moreira 787, 88704-900 Tubar\~{a}o, S.C.,
Brazil}
\begin{abstract}

We use the optimized perturbation theory, or linear $\delta$
expansion, to evaluate the critical exponents in the critical $3d$
O($N$) invariant scalar field model. Regarding the implementation
procedure, this is the first successful attempt to use the method
in this type of evaluation. We present and discuss all the
associated subtleties producing a prescription which can, in
principle, be extended to higher orders in a consistent way.
Numerically, our approach, taken at the lowest nontrivial order
(second order) in the $\delta$ expansion  produces a modest
improvement in comparison to mean field values for the anomalous
dimension $\eta$ and correlation length $\nu$ critical exponents.
However, it nevertheless points to the right direction of the
values obtained with other methods, like the $\epsilon$-expansion.
We discuss the possibilities of improving over our lowest order
results and on the convergence to the known values when extending
the method to higher orders.

\end{abstract}

\pacs{05.70.Fh, 11.10.Wx, 64.60.Fr}

\maketitle

\centerline{\it Version with some corrections in relation to the published one in:
Physica {\bf A342}, 570 (2004).}

\section{INTRODUCTION}

A distinct feature as we approach second order phase transition points
is the emergence of critical phenomena and associated universality and
scaling properties, as a result of diverging correlation lengths. Thanks
to these characteristics we can relate many different systems just by
simple general distinctions, like dimension and symmetry, with their
universal behavior set by critical exponents, independent of the
microscopic dynamics \cite{critexp}. On the other hand, close to the
critical temperature of transition we are faced with the problem of
breakdown of perturbation methods in field theory due to the appearance
of infrared divergences, as the correlation lengths diverge (or masses
vanish), which then require the use of nonperturbative methods to study
the physical system around the critical points and, ultimately, the
phase transition process itself. The present work was motivated by the
recent progress in dealing with field theory phase transitions in the
context of the nonperturbative method of the optimized mass, or linear
$\delta$ expansion (LDE) \cite{prb,pra,prl,pra03}. Here, for the first
time, this method is used in the derivation of the critical exponents of
the three-dimensional O($N$) invariant $\lambda \phi^4$ model.

Critical exponents have been extensively studied at different levels of
accuracy, particularly within the $\epsilon$-expansion and
renormalization group (RG) methods (see, e.g. Ref. \cite{justin} for a
throughout discussion of various methods of evaluating critical
exponents in field theory). However, it is well known that these methods
may not be applicable or appropriate for a number of important physical
systems. Examples include, for instance, the reliability of the
$\epsilon$-expansion in the study of the electroweak phase transition,
where there may not be a visible fixed point in three dimensions when we
perform an expansion around dimension $d=4-\epsilon$ \cite{russell}.
There are also the important cases of multicritical phenomena in two
dimensions with no obvious upper critical dimension or with no O($N$)
invariants, in which case nonperturbative techniques like the
$\epsilon$-expansion
and large-$N$ approximation break down \cite{multi}. The use of
renormalization
group techniques have shown recently to produce
impressive good results to
critical exponents \cite{wett}, however,
there we commonly have to deal with a
set of flow equations that may
become quickly very complicate to solve for
complex systems.

Using the LDE  one makes use of an interpolation of the original
action by modifying the masses and coupling constants in terms of
a fictitious parameter $\delta$ that at the end is taken to the
unity, so as to recover the original model. At the same time an
arbitrary mass parameter must be introduced to balance the
dimensions of the interpolated theory. At the end it is fixed in
such a way that nonperturbative information can be taken into
account in quantities which have been perturbatively computed.
Concerning the O($N$) invariant scalar field model in three
dimensions, the convergence properties of the LDE have been
studied recently in great detail, specifically at the critical
point of phase transition in the large-$N$ limit
\cite{prl,pra03,braatenradescu}. Those results are also
encouraging concerning the computation of critical exponents as
performed in this paper. An advantage of the LDE method when
compared to other nonperturbative methods is that we remain all
the time within the familiar grounds of perturbation theory. The
method then becomes of much simpler use than the traditional ones
and we expect that the LDE may even further improve the accuracy
of other nonperturbative methods when used in conjunction with
them. We also expect the method to be suitable in the evaluation
of the critical properties in the transition point in those cases
where no clear expansion parameter exists or when the conventional
techniques do not apply, as in the models discussed in the
previous paragraph. The LDE has been successfully applied before
to study field theory at finite temperature and phase transitions
in those situations where conventional perturbation theory breaks
down \cite{prb,pra,prd1}. In particular, in Ref. \cite{prd2}, the
LDE was applied to study the phase transition patterns in coupled
scalar field models reproducing previous renormalization group
applications to the problem and also predicting new critical
points, showing the advantages of the method for applications
regarding complex systems that may exhibit multicritical points.

In this work we shall consider the same O($N$) model studied in
Refs. \cite{prb,pra,prl,pra03}. It is described by
the  O($N$) invariant super-renormalizable
action

\begin{equation}
S_{\phi}=  \int d^3x \left [ \frac {1}{2} | \nabla \phi |^2 +
\frac {1}{2} r_0
\phi^2 + \frac {u}{4!} (\phi^2)^2
\right ] \;,
\label{action2}
\end{equation}
where $r_0 = r + A$, with $A$ representing the mass renormalization
counterterm needed to remove any ultraviolet divergence. The full
propagator for this theory is given by

\begin{equation}
G(p)= \left[ p^2  + m^2
+ \Sigma_{\rm ren}(p) - \Sigma_{\rm ren}(0) \right  ]^{-1}\;,
\label {originalG}
\end{equation}
where $m^2 = r + \Sigma_{\rm ren}(0)$ with $\Sigma_{\rm ren}$
representing the renormalized self-energies. At the critical
point, $m^2=0$, which implies the Hugenholtz-Pines theorem, $r_c=
- \Sigma_{\rm ren}(0)$. Using the model described by Eq. (\ref
{action2}) is advantageous since our results can readily be
compared with other results that have been extensively obtained in
connection with the $\phi^4$ O($N$) invariant scalar model. This
is a particularly important model also for the reason that it can
describe the critical properties (due to universality) of many
different physical systems, like polymers ($N=0$), Ising models
($N=1$), superfluid Helium, Bose-Einstein condensation of atomic
atoms ($N=2$), ferromagnets ($N=3$), the scalar sector of the
electroweak standard model ($N=4$), etc. A similar attempt to
compute critical exponents with the $\delta$ expansion was
performed by the authors of Ref. \cite{gandhi}. However, they made
use of the nonlinear version of method which leads to a
considerable more complicated perturbation series (see also
\cite{bjp}), which quickly becomes cumbersome beyond leading order
in $\delta$ and their results were not so good for the same
critical exponents evaluated here. On the other hand, the way the
linear version is employed here avoids those difficulties and the
results can in principle be extended to arbitrary orders as usual
perturbative calculations. Our results for the correlation length
(or mass) critical exponent $\nu$ and the anomalous dimension
exponent $\eta$, already to the lowest nontrivial order (second
order) in $\delta$, though showing only modest improvements over
the mean field values, are shifted away from these in the
direction of the values known from high precision numerical
results despite the simplicity of our calculations.

This paper is organized as follows. In Sec. II we briefly review the LDE
method and present the interpolated version of the O($N$) invariant
scalar field action relevant to our study. In Sec. III, we carry out
the formal evaluation of the critical exponents for the correlation
length $\nu$ and anomalous dimension $\eta$. Our concluding remarks are
presented in Sec. IV, where we also discuss the extension of our
method to higher orders.

\section {LDE AND THE INTERPOLATED EFFECTIVE SCALAR MODEL}

Let us start our work by reviewing the application of the LDE method to
our problem. The LDE was conceived to treat nonperturbative physics
while staying within the familiar calculation framework provided by
perturbation theory. In practice, this can be achieved as follows.
Starting from an action $S$ one performs the following interpolation

\begin{equation}
S \rightarrow S_{\delta} =  \delta S + (1 - \delta) S_0(M) \;,
\label{s}
\end{equation}
where $S_0$ is the soluble quadratic action, added by an (optimizable)
mass term $M$, and $\delta$ is an arbitrary parameter. The above
modification of the original action somewhat reminds the usual trick
consisting of adding and subtracting a mass term to the original action.
One can readily see that at $\delta=1$ the original theory is retrieved,
so that $\delta$ actually works just as a bookkeeping parameter. The
important modification is encoded in the field dependent quadratic term
$S_0(M)$ that, for dimensional reasons, must include terms with mass
dimensions ($M$). In principle, one is free to choose these mass terms
and within the Hartree approximation they are replaced by a direct (or
tadpole) type of self-energy before one performs any calculation. In the
LDE they are taken as being completely arbitrary mass parameters, which
will be fixed at the very end of a particular evaluation by an
optimization method. One then formally pretends that $\delta$ labels
interactions so that $S_0$ is absorbed in the propagator whereas $\delta
S_0$ is regarded as a quadratic interaction. So, one sees that the
physical essence of the method is the traditional dressing of the
propagator to be used in the evaluation of physical quantities very much
like in the Hartree case. What is different between the two methods is
that, within the LDE the propagator is completely arbitrary, whereas it
is constrained to cope only with the so-called direct terms ({\it i.e.}
tadpoles) within the Hartree approximation. So, within the latter
approximation the relevant contributions are selected according to their
topology from the start.

Within the LDE one calculates in powers of $\delta$ as if it was a small
parameter. In this respect the LDE resembles the large-$N$ calculation
since both methods use a bookkeeping parameter which is not a physical
parameter like the original coupling constants and within each method
one performs the calculations formally working as if $N \rightarrow
\infty$ or $\delta \rightarrow 0$, respectively. {}Finally, in both
cases the bookkeeping parameters are set to their original values at the
end which, in our case, means $\delta=1$. However, quantities evaluated
at any finite LDE order from the dressed propagator will depend
explicitly on $M$, unless one could perform a calculation to all orders.
Up to this stage the results remain strictly perturbative and very
similar to the ones which would be obtained via a true perturbative
calculation. It is now that the freedom in fixing $M$ generates
nonperturbative results. Since $M$ does not belong to the original
theory one may require that a physical quantity $\Phi^{(k)}$ calculated
perturbatively to order-$\delta^k$ be evaluated at the point where it is
less sensitive to this parameter. This criterion, known as the Principle
of Minimal Sensitivity (PMS), translates into the variational relation
\cite{pms}

\begin{equation}
\frac {d \Phi^{(k)}}{d M}\Big |_{{\overline M}, \delta=1} = 0 \;.
\label{pms}
\end{equation}

The optimum value $\overline M$ which satisfies Eq. (\ref{pms}) must be
a function of the original parameters including the couplings, which
generates the nonperturbative results.

{}Following the considerations above, in order to interpolate the original
 theory
described by Eq. (\ref {action2}), one may use

\begin{equation}
S_0 =  \frac{1}{2}  \left [ | \nabla \phi|^2 + r\phi^2+ M^2 \phi^2  \right ] \;,
\label{S0}
\end{equation}
as in Refs. \cite {pra}.
Then, the interpolated action reads

\begin{equation}
S_{\delta}=  \int d^3x \left [ \frac {1}{2} | \nabla \phi |^2 +
\frac {1}{2} (r +M^2) \phi^2  -\frac{\delta}{2}M^2 \phi^2 +
\frac {\delta u}{4!} (\phi^2)^2   + \frac {\delta}{2} A_{\delta} \phi^2
\right ] \;\;,
\label{Sdelta}
\end{equation}

\noindent
where $A_{\delta}$ represents the renormalization mass
counterterm for the interpolated theory, which depends on the parameters
$M$ and $\delta$. It is important to note that by introducing only extra
mass terms in the original theory the LDE does not alter the polynomial
structure and, hence, the renormalizability of a quantum field theory.
In practice, the original counterterms change in an almost trivial way
so as to absorb the new $M$ and $\delta$ dependence (for details, see
for instance \cite{pra}). The compatibility
of the LDE with the renormalization program has been shown in the
framework of the O($N$) scalar field theory at finite temperatures, in
Ref. \cite{prd1}, showing that it consistently takes into account
anomalous dimensions in the critical regime.

{}For the interpolated theory, the full propagator $G^{(\delta)} (p)$,
can be written as

\begin{equation}
G^{(\delta)}(p)= \left[ p^2  +   m^2 +(1-\delta)M^2
+ \Sigma^{(\delta)}_{\rm ren}(p) - \Sigma^{(\delta)}_{\rm ren}(0)\right
]^{-1}\;, \label {G1}
\end{equation}
where $\Sigma^{(\delta)}_{\rm ren}$ are the (renormalized) momentum
dependent self-energies evaluated in powers of $\delta$ and $m^2 = r +
\Sigma^{(\delta)}_{\rm ren}(0)$. At the same time, the bare propagator
is given by

\begin{equation}
G^{(0)}(p)=[p^{2}+\Omega^{2}]^{-1} \,\,,
\label{propafora}
\end{equation}
where $\Omega^{2}= m^2 + M^2$. It is interesting to note that at the
critical point ($m=0$) the propagator given by Eq. (\ref {propafora})
does not generate any infrared divergences since it is automatically
regulated by the LDE mass parameter $M$ whereas the equivalent bare
propagator of the {\it original} theory is massless. In fact, this is
the main problem concerning any eventual perturbative evaluation of
quantities at the critical point, an example being the problem of
calculations of critical temperature shifts, $\Delta T_c$
\cite{prb,pra,prl,pra03}.

In our evaluation of the renormalized quantities and fixed points we
shall use the following {}Feynman rules for the vertices: $\delta M^{2}$,
$-\delta A_{\delta}$ for the quadratic interaction and $-\delta u$ for
the quartic one. Those rules are used to evaluate, {\it perturbatively}
in $\delta$, all the relevant physical quantities. After that, as
discussed in Sec. II, the nonperturbative results are produced by means
of the PMS condition, Eq. (\ref{pms}).

\section{ CRITICAL EXPONENTS EVALUATION UP TO ORDER-$\delta^2$}

We now proceed to the evaluation of the critical exponents for the
correlation length and anomalous dimension, which follow from the
usual scaling relations at the critical point. {}Following the standard
conventions and definitions given, e.g., in Parisi's textbook
\cite{parisi} one writes the scaling relations as

\begin{equation}
m^{2}\propto(r-\delta r_{c}^{\delta})^{2} \Rightarrow
Z_{2} \equiv\frac{dr}{dm^{2}}\sim Z_1 m^{2C_{2}} \Rightarrow
C_{2}=\frac{1}{2\nu}-1  \;,
\label{mscale}
\end{equation}
and

\begin{equation}
Z_{1}\equiv \left\{ \frac{\left[G^{(\delta)}(p) \right]^{2}}{
\frac{dG^{(\delta)}(p)}{dp^{2}}} \right\}\Bigr|_{p^{2}=0}\propto m^{2C_{1}}
\Rightarrow C_{1}=\frac{\eta}{2} \;,
\label{Zscale}
\end{equation}

\noindent
where $Z_{1}$ is the field renormalization function and
$\lambda$ is the effective renormalized coupling, defined in
terms of the one-particle irreducible four-point function
$\Gamma_{4}(0)$ as

\begin{equation}
\lambda=Z_{1}^{2}\frac{\Gamma_{4}(0)}{m} \;.
\label{lambda}
\end{equation}

In order to compute the critical exponents defined in Eqs.
(\ref{mscale}) and (\ref{Zscale}), we need to compute the
$\beta$-function,

\begin{equation}
\beta=m^{2}\frac{\partial}{\partial m^{2}}\lambda =
-\frac{g}{2}\frac{\partial}{\partial g}\lambda\;,
\label{beta}
\end{equation}
from which the fixed points are obtained. Note that $g=u/m$ is taken as
a dimensionless coupling. The fixed points are then obtained, as usual,
from the solutions of $\beta(\lambda_{c})=0$.

{}From the relations (\ref{mscale}) and (\ref{Zscale}) we can also define

\begin{equation}
C_{1}(\lambda)\equiv -\frac{g}{2}\frac{\partial}{\partial g}
\ln[Z_{1}(\lambda)]\;,
\label{C1}
\end{equation}
and

\begin{equation}
C_{2}(\lambda)\equiv -\frac{g}{2}\frac{\partial}{\partial g}
\ln[Z_{2}(\lambda)/Z_1(\lambda)]\;,
\label{C2}
\end{equation}
which, when evaluated at the fixed point, will determine the
constants $C_{1}=C_{1}(\lambda_{c})$ e $C_{2}=C_{2}(\lambda_{c})$
defining the critical exponents in (\ref{mscale}) and
(\ref{Zscale}).

Having introduced the quantities given above, let us now use the LDE to
evaluate $\Gamma_4(0)$, $Z_1$ and the self-energy $\Sigma_{\rm
ren}^{(\delta)}$ expanded in powers of $\delta$. Note that at first
order in $\delta$ the results are trivial, since it corresponds just to
the lowest order one-loop perturbative expansion for which $Z_1 = Z_2
=1$. In this case, using the definitions (\ref{C1}) and (\ref{C2}), one
obtains $C_1=C_2=0$ which lead to the well known mean-field critical
exponents $\eta=0$ and $\nu=1/2$. The next order results in $\delta$
will however already lead to results departing from the mean-field ones.
{}For consistency we carry the calculations of all quantities to the
first non-trivial order, $\delta^2$. Also, all our calculations are
performed using dimensional regularization and the modified minimal
subtraction (${\overline {\rm MS}}$) renormalization scheme which amounts
to replace the the momentum integrals by

\begin{equation}
\int \frac {d^3 p}{(2 \pi)^3} \to \int_p
\equiv \left(\frac{e^{\gamma_E} \mu_s^2}{4 \pi} \right)^\epsilon
\int \frac {d^{2\omega} p}{(2 \pi)^{2 \omega}} \;,
\label{rdim}
\end{equation}

\noindent
where $2\omega=3-2\epsilon$ while $\mu_s$ is an arbitrary mass scale and
$\gamma_E \simeq 0.5772$ is the
Euler-Mascheroni constant.

\subsection{Evaluating the optimized self-energy.}

As already emphasized, all relevant physical quantities evaluated
perturbatively with the LDE will depend on the arbitrary quantity
$\Omega^2= m^2 + M^2$, which is present in the LDE bare propagator, Eq.
(\ref{propafora}). Nonperturbative results can be generated by
optimizing the renormalized self-energy, $\Sigma_{\rm
ren}^{(\delta)}(0)$, which then generates the optimum ${\overline M}$.
Having mass dimensions, this quantity can be expressed as a function of
the original $m$ and $u$ parameters which, as we shall see, will allow
us to express the optimum quantity ${\overline \Omega}$ in terms of $m$
and the dimensionless coupling $g=u/m$. The choice of $\Sigma_{\rm
ren}^{(\delta)}(0)$ as the physical quantity to be optimized is well
justified since this is an important quantity in the study of critical
phenomena signaling phase transitions via the Hugenholtz-Pines theorem.
It also fixes the effective scale determining all critical exponents through
the scaling relations at the critical point.
To order-$\delta^2$ the self-energy has been explicitly evaluated in
Ref. \cite{pra}. When $m \ne 0$ it is given by \footnote{See Ref. \cite
{pra} for details concerning the $m=0$ case.}

\begin{eqnarray}
\Sigma_{\rm ren}^{(2)}(0) &=& -\frac{\delta u
\Omega}{8\pi}\left(\frac{N+2}{3}\right) +
\frac{\delta^{2}u M^2}{16\pi\Omega}
\left(\frac{N+2}{3}\right) + \frac{\delta^{2}u^{2}}{128\pi^2}
\left(\frac{N+2}{3}\right)^{2}  \nonumber\\
&-&\frac{\delta^{2}u^{2}}{(8\pi)^{2}} \frac{(N+2)}{18}
\left[4\ln\left(\frac{\mu_s}{2 \Omega}\right) + 2 +
4 \ln\left(\frac{2}{3}\right) \right] +O(\delta^3) \;.
\label{Sigma0}
\end{eqnarray}
We then set $\delta=1$ solving the PMS equation

\begin{equation}
\frac {d \Sigma^{(2)}_{\rm ren}(0)}{d M}\Big |_{{\overline M}} = 0 \;,
\label{pmsM}
\end{equation}
to obtain the ($N$-independent) roots

\begin{eqnarray}
&& {\overline M}_0=0 \nonumber \\
&& {\overline M}_\pm^2= \frac{2}{(12 \pi)^2}\left\{u^2 \pm u\left[(12 \pi m)^2 +
u^2 \right]^{1/2} \right\} \;.
\label{Mbar}
\end{eqnarray}
The solutions ${\overline M}_0$ and ${\overline M}_{-}$, at the
critical point $m = 0$, are trivial ones, while ${\overline
M}_{+}(m\to 0)=u/(6 \pi)$ remains nonzero even at the critical
point and it is then able to effectively lead to nonperturbative
results as shown below, apart from   agreeing with our previous
results \cite{pra}. This nontrivial root (omitting the index ``+"
from now on) can be written in terms of the dimensionless
coupling, $g=u/m$, as

\begin{equation}
{\overline M}^2= \frac {m^2 }{72 \pi^2} \left\{ g^2 + 12 \pi g \left[ 1 +
\frac {g^2}{(12 \pi)^2} \right]^{1/2} \right\} \;\;,
\label{Mbarg}
\end{equation}
which is, manifestly, a nonperturbative quantity.
{}Finally, the
optimized ${\overline \Omega}^2 = m^2 +{\overline M}^2$ can be
written as

\begin{equation}
{\overline \Omega}^2 = m^2 \left \{ 1 + \frac{1}{72 \pi^2} \left [ g^2 + 12
\pi g F(g) \right ] \right \} \;\;,
\label{Obar}
\end{equation}
where we have defined

\begin{equation}
F(g)=\left [ 1 +
\frac {g^2}{(12 \pi)^2} \right ]^{1/2} \;\;.
\label {f(g)}
\end{equation}
So, in the following, our strategy will be to evaluate all relevant
quantities with the LDE to order-$\delta^2$ and then perform the
substitution $\Omega \to {\overline \Omega}$ where the latter quantity
is given by Eq. (\ref {Obar}).

\subsection{Evaluating the critical coupling $g_c$}

Let us start by evaluating the $\beta$ function whose roots define
$g_c$. The relevant Green's function is the four-point one, which
to order-$\delta^2$ reads

\begin{eqnarray}
\Gamma_{4}(0)&=&\delta u - \frac{3}{2}\delta^{2}u^{2}
\left(\frac{N+8}{9}\right) \int \frac{d^{3}q}{(2\pi)^{3}}
\frac{1}{(q^{2}+\Omega^{2})^{2}} \nonumber\\ &=& \delta u -
\frac{3\delta^{2}u^{2}}{16\pi\Omega}\left( \frac{N+8}{9} \right)\;,
\label{Gamma4}
\end{eqnarray}
where the first term on the RHS  is the tree vertex and the second
term is the one-loop correction. In addition to the four-point
function, according to Eq. (\ref {lambda}), one needs the field
renormalization function in order to define the effective
renormalized coupling. To evaluate  $Z_1$, as given by Eq.
(\ref{Zscale}), we need the full propagator $G^{(\delta)}(p)$
which, to order-$\delta^2$, is

\begin{equation}
G^{(2)}(p)=[p^{2}+m^{2}+\Sigma^{(2)}_{\rm ren}(p)
-\Sigma_{\rm ren}^{(2)}(0)]^{-1} \,\,,
\label{G2}
\end{equation}
where $\Sigma_{\rm ren}^{(2)}(p)-\Sigma_{\rm ren}^{(2)}(0)$ is
given by \cite{braatenradescu}

\begin{equation}
\Sigma_{\rm ren}^{(2)}(p)-\Sigma_{\rm ren}^{(2)}(0)=
-\frac{(N+2)\delta^{2}u^{2}}{18(4\pi)^{2}} \left [1
-\frac{3\Omega}{p}\arctan\left(\frac{p}{3\Omega}\right)
-\frac{1}{2}\ln
\left(\frac{p^{2}+9\Omega^{2}}{9\Omega^{2}}\right) \right ] \;.
\label{braaten}
\end{equation}

Substituting Eqs. (\ref{braaten}) and (\ref{G2}) in (\ref{Zscale}),
one gets

\begin{equation}
Z_{1}=\left\{
1+\frac{(N+2)\delta^{2}u^{2}}{18(4\pi)^{2}54\Omega^{2}}
\right\}^{-1} \;\;\;,
\label{Z1}
\end{equation}
which, after expanding  to order-$\delta^2$, becomes

\begin{equation}
Z_{1}=1- \frac{(N+2)\delta^{2}u^{2}}{18(4\pi)^{2}54\Omega^{2}}\;.
\label{Z12}
\end{equation}

{}From Eqs. (\ref{lambda}), (\ref{Gamma4}) and (\ref{Z12}) we then
obtain the renormalized coupling to order $\delta^2$

\begin{equation}
\lambda = \frac{\delta u}{m} - \frac{3\delta^{2}u^{2}}{16\pi m \Omega}
\left(\frac{N+8}{9}\right) + O(\delta^{3})\;\;,
\label{lambdaM}
\end{equation}
from which the $\beta$-function will follow via Eq. (\ref{beta}).
But before performing the derivatives with respect to $g$ (or $m$)
let us recall that this quantity also appears in ${\overline
\Omega}$. Then, substituting the optimized quantity Eq.
(\ref{Obar}) in (\ref{lambdaM}) and setting $\delta=1$ we  obtain
the optimized renormalized coupling given by

\begin{equation}
{\overline \lambda} (g)=  g - \frac{g^{2} (N+8)}{48\pi}
\left \{ 1 + \frac{1}{72 \pi^2} \left [ g^2 + 12
\pi g F(g) \right ] \right \}^{-1/2}  \;,
\label{g1}
\end{equation}
which is, of course, a nonperturbative quantity. Now, by using
the definition for the $\beta$-function, which from Eq. (\ref{g1})
can be expressed as

\begin{equation}
\beta =
-\frac{g}{2}\frac{\partial}{\partial g}{\overline \lambda}\;,
\label{betaop}
\end{equation}
one finds the fixed points, $g_c$, as given by the solutions of
$\beta(g_c)=0$. {}For a fixed value of $N \neq 0$ we always find
three roots, a trivial one ($g_c=0$), a purely imaginary one
(which leads to unacceptable scaling relations at the critical point)
and a positive real root. This last one can easily be found
numerically, and is given, e.g. for a few cases, by
$g_c(N=1) = 15.3524$, $g_c(N=2) = 12.2968$, $g_c(N=3) = 10.3692$.
The nontrivial real positive roots are the ones  that will be used in
the subsequent determination of the critical exponents.

\subsection{Evaluating the critical exponents}

Now, the evaluation of the critical exponent $\eta$ is fairly easy.
{}From Eqs. (\ref{Zscale}) and (\ref{C1}) we have that

\begin{equation}
\eta = - g \frac {\partial}{\partial g} \ln [{\overline
Z_1}(g)]\Bigr|_{g=g_c} \;, \label{etaderiva}
\end{equation}
where ${\overline Z_1}$ represents the optimized field renormalization
function. This quantity can be obtained directly from Eq. (\ref {Z12})
with the replacement $\Omega \to {\overline \Omega}$. Then, setting
$\delta=1$, one obtains

\begin{equation}
{\overline Z_{1}}=1- \frac{g^{2} (N+2)}{18(4\pi)^{2}54}
\left \{ 1 + \frac{1}{72 \pi^2} \left [ g^2 + 12
\pi g F(g) \right ] \right \}^{-1}\;.
\label{Z1otim}
\end{equation}
{}From Eq. (\ref{etaderiva}) and the previous results for the fixed
points, we then obtain the results (for some representative values of
$N$) $\eta(N=1) = 0.0026$, $\eta(N=2) = 0.0029$, $\eta(N=3) = 0.0030$.
These values are contrasted to the results from other methods in Table I.
Results for other values of $N$ can also be easily obtained from the
previous equations.

We now turn to the calculation of the critical exponent $\nu$ to
order-$\delta^2$. According to our prescription, to evaluate this
quantity one computes the optimized ${\overline Z_2}$ which is
given by

\begin{equation}
{\overline Z_2} = \frac{d {\overline r}}{d m^2}\;,
\end{equation}
where ${\overline r} = m^2 - {\overline \Sigma}^{(2)}_{\rm ren}(0)$. The
optimized self-energy at order $\delta^2$,
${\overline \Sigma}^{(2)}_{\rm ren}(0)$, can be
trivially obtained from Eq. (\ref{Sigma0}) by performing the
substitutions $M \to {\overline M}$ and $\Omega \to {\overline \Omega}$
as given by Eqs. (\ref {Mbarg}) and (\ref {Obar}). The critical
exponent $\nu$ then follows from Eqs. (\ref{C2}) and (\ref{mscale}),
or

\begin{equation}
\nu = \left \{ 2 - g \frac {\partial}{\partial g}
\ln [{\overline{Z}_2(g)/\overline{Z}_1(g)}] \right \}^{-1}\Bigr|_{g=g_c} \;,
\end{equation}
which, at $\delta=1$ and for a few representative values of $N$, yields
the results $\nu(N=1)=0.5287$, $\nu(N=2)=0.5362$, $\nu(N=3)=0.5422$.
These results,
together with those for $\eta$, are shown in Table I so that they can
be compared with the results obtained using different approximations.

The $\epsilon$-expansion results for the critical exponents can be found
e.g. in Ref. \cite{justin}. In Parisi's book \cite{parisi} one can find,
for instance the results for the critical exponents for one and two-loop
perturbation theory (PT). In Ref. \cite{kleinert1}, it is used a
variational perturbation theory (VPT) to continue the renormalization
constants of three-dimensional $\phi^{4}$ theories to the regime of
strong bare coupling. The authors of Ref. \cite{gandhi} have used
the logarithmic $\delta$-expansion (DE-log) to
obtain the exponents $\eta$ and $\nu$. All the results from
these different methods are shown in
Table I together with ours.


\begin{table}
\begin{center}
\caption{Numerical results for $\nu$ and $\eta$.}
\begin{tabular}{|c|cc|cc|}\hline
                        & & (N=1)& & (N=1)\\
${\rm Method}$          & $\nu$  &(N=2)          & $\eta$ &(N=2) \\
                         & & (N=3)  & & (N=3) \\\hline
                        & $0.5287$ &     &  $0.0026$ & \\
${\rm LDE}-O(\delta^2)$ & $0.5362$  &  & $0.0029$ & \\
                        & $0.5422$       & & $0.0030$  &     \\ \hline
                        & $0.626$          &   &  $0.037$  &     \\
$\epsilon-{\rm expansion}$    & $0.655$  &  & $0.039$ & \\
                        & $0.679$        &    & $0.039$   &        \\ \hline
                        & $0.6$    &    & $0$   &     \\
${\rm PT}\;{\rm one-loop}$      & $0.625$  &   & $0$  & \\
                        & $0.647$  &     & $0$  &       \\ \hline
                        & $0.630\pm 0.015$    &     & $0.031\pm 0.004$ &  \\
${\rm PT}\;{\rm two-loops}$       & $--$      & &  $--$     & \\
                        & $--$      &    & $--$   &         \\ \hline
                        & $0.63$       &    & $0.030$     &    \\
${\rm VPT}$             & $0.670$ & & $0.032$ & \\
                        & $0.706$         &  & $0.032$      &  \\ \hline
${\rm DE}-{\rm log}$          & $0.53$& &$0.0013$ &\\
& $--$      & &  $--$     & \\
& $--$      &    & $--$   &         \\ \hline
\end{tabular}
\end{center}
\end{table}

\section{CONCLUSIONS}

In this paper, the LDE has been applied for the first time
in the evaluation of critical exponents. An early attempt to
use this type of method was carried out by Gandhi and McKane
\cite{gandhi} who used a variant of the method, known as the logarithmic
$\delta$ expansion. However, the authors did not implement the method
correctly, since in their implementation they have not made use of any
optimization scheme, just plain expansion in $\delta$, which led to many
difficulties associated to the fact that their
fixed point coupling $g_c$ behaves like $1/\delta$.

Here, our main concern was to show how to circumvent the
difficulties found in Ref. \cite{gandhi}. Our approach to the
problem is based on the central idea of first obtaining optimized
values for the arbitrary parameter $M$ by applying the
optimization procedure PMS to the renormalized self-energy at a
given order in $\delta$, $\Sigma_{\rm ren}^{(\delta)}(0)$, which
then generates an optimum value for $M$ in terms of the original
parameters of the theory ($m$ and $g=u/m$). Then, the full
propagator and four-point functions are evaluated to the same
perturbative order with the LDE and their optimum values obtained
by performing the substitution $M \to {\overline M}(m,g)$. Our
analytical results show explicitly that the ${\overline M}$
dependence on $g$ is nonperturbative (all orders are present). The
relevant optimized quantities ${\overline Z_1}$, ${\overline Z_2}$
and the ${\overline \beta}$ function are obtained upon using
derivatives with respect to $g$. This whole procedure then
generates the optimized values for the fixed point, $g_c$, as well
as for the critical exponents $\eta$ and $\nu$ (note that, in
contrast to the approach used in \cite{gandhi}, the optimization
procedure is already performed prior to computing the
$\beta$-function and the fixed points). In principle this
prescription can be extended to any perturbative order in
$\delta$. The only difficulties of extending our results to higher
orders are technical ones, associated to the evaluation of
multi-loop diagrams containing massive propagators.

As far as our numerical values are concerned, they already at
nontrivial lowest order show improvement, though modest, over the
mean field theory ones. In a sense, the physical quantities
evaluated by us with the LDE to order-$\delta^2$ contain only
two-loop diagrams so it comes as no surprise the fact that our
lowest order results are not as good as the ones obtained by
resumming higher order corrections. One expects, as shown by the
many LDE applications, that by considering higher order
contributions one can quickly obtain better results from a
calculation which has the advantage of being completely
perturbative regarding the evaluation of {}Feynman diagrams. In
this context, from our previous experience \cite{pra,prl,pra03} in
extending to higher order the LDE to obtain the a the critical
temperature shift for an homogeneous Bose gas, which is a problem
that share many similarities to the one studied here, we can
advance a number of important issues that must be handled. {}First
is the slow convergence behavior observed in \cite{braatenradescu}
by the direct extension of the LDE method studied here. Since both
the calculation of the temperature shift for Bose condensation
(for $N=2$) and the exponents computed here make use of the same
critical model, it is reasonable to expect that the same
convergence behavior will show up  in higher order calculations of
the critical exponents within the LDE method. The general
explanation and throughout understanding of the slow convergence
behavior for the Bose condensate critical temperature shift was
given in Ref. \cite{pra03}. Also, in the same reference, it was
discussed how appropriate resummation techniques can speed
convergence. The same resummation techniques used there could also
be used here  in order to improve convergence. A second important
issue that commonly arises when extending the LDE beyond second
order is the appearance of imaginary solutions upon the use of the
optimization, Eq. (\ref{pms}). In previous references we have
shown that only the real part of those solutions make physical
sense and indeed they lead to correct results, despite some
obvious embarrassment of having to deal with those imaginary
solutions. To circumvent this problem an alternative optimization
has been proposed recently \cite{ipms}, which make use of
additional parameters from third order onwards. In practice this
can be achieved by replacing, in Eq. (\ref{s}),  terms like
${M^*}^2=(1-\delta)M^2$ by terms like

\begin{eqnarray}
&&M^* =\left[1-a_1\delta-(1-a_1)\delta^2\right]^{1/2} M, \;\; {\rm
at\; third \; order}
\nonumber \\
&&M^* = \left[1-a_1\delta-(1-a_1)\delta^2+a_2^2 \delta^2 (1-\delta)
\right]^{1/2} M, \;\; {\rm at\; fourth\; order} \;,
\end{eqnarray}
and so on, where the $a_j$, $j=1,\ldots,n-2$ are additional parameters 
added at order $n$ in the LDE and to be determined with $M$ by generalizing 
the optimization condition Eq. (\ref{pms}) to a system of equations,

\begin{eqnarray}
&&\frac {d \Phi^{(n)}}{d M}\Big |_{{\overline M},\bar{a}_1,\ldots,\bar{a}_{n-2},
\delta=1}
= 0 \;,\nonumber \\
&&\frac {d^2 \Phi^{(n)}}{d M^2}\Big |_{{\overline M},\bar{a}_1,\ldots,\bar{a}_{n-2}, 
\delta=1}
= 0 \;,\nonumber \\
&& .\nonumber \\
&& .\nonumber \\
&& .\nonumber \\
&&\frac {d^{n-1} \Phi^{(n)}}{d M^{n-1}}\Big |_{{\overline M},\bar{a},\ldots,\bar{a}_{n-2}, 
\delta=1}
= 0 \;.
\end{eqnarray}
As shown in \cite{ipms} this optimization procedure generates, in
the Bose-Einstein condensation case,  only real values for the LDE
interpolating  parameters with the added bonus of fast convergence
already at lowest orders. As far the LDE application of Ref. \cite{ipms} to the 
Bose-Einstein condensation problem
is concerned, it has produced some of the most precise and stable
analytical predictions for the critical temperature shift.
Those results thus also give an indication of the applicability of the use
of the LDE in analogous models close and at the critical point. 
We are currently working on the use of this modified optimization
to the problem studied in this paper and we will report
the results elsewhere.

\acknowledgments

M.B.P. and R.O.R. were partially supported by Conselho Nacional de
Desenvolvimento Cient\'{\i}fico e Tecnol\'ogico (CNPq-Brazil).
P.J.S. were partially supported by Associa\c{c}\~{a}o Catarinense
das Funda\c{c}\~{o}es Educacionais (ACAFE).

\end{document}